\documentclass[twocolumn,aps,pra,showpacs]{revtex4}
\usepackage[T1]{fontenc}
\usepackage[latin9]{inputenc}
\usepackage{mathrsfs}
\usepackage{amsmath}
\usepackage{amssymb}
\usepackage{graphicx}
\usepackage{esint}

\makeatletter
\@ifundefined{textcolor}{}
{%
 \definecolor{BLACK}{gray}{0}
 \definecolor{WHITE}{gray}{1}
 \definecolor{RED}{rgb}{1,0,0}
 \definecolor{GREEN}{rgb}{0,1,0}
 \definecolor{BLUE}{rgb}{0,0,1}
 \definecolor{CYAN}{cmyk}{1,0,0,0}
 \definecolor{MAGENTA}{cmyk}{0,1,0,0}
 \definecolor{YELLOW}{cmyk}{0,0,1,0}
}

\makeatother

\begin{document}

\title{Equation of state and contact of a strongly interacting Bose gas
in the normal state}

\author{Xia-Ji Liu$^{1}$, Brendan Mulkerin$^{1}$, Lianyi He$^{2}$, and
Hui Hu$^{1}$}

\affiliation{$^{1}$Centre for Quantum and Optical Science, Swinburne University
of Technology, Melbourne 3122, Australia}

\affiliation{$^{2}$Theoretical Division, Los Alamos National Laboratory, Los
Alamos, NM 87545, USA}

\date{\today}
\begin{abstract}
We theoretically investigate the equation of state and Tan's contact
of a non-degenerate three dimensional Bose gas near a broad Feshbach
resonance, within the framework of large-$N$ expansion. Our results
agree with the path-integral Monte Carlo simulations in the weak-coupling
limit and recover the second-order virial expansion predictions at
strong interactions and high temperatures. At resonance, we find that
the chemical potential and energy are significantly enhanced by the
strong repulsion, while the entropy does not change significantly.
With increasing temperature, the two-body contact initially increases
and then decreases like $T^{-1}$ at large temperature, and therefore
exhibits a peak structure at about $4T_{c0}$, where $T_{c0}$ is
the Bose-Einstein condensation temperature of an ideal, non-interacting
Bose gas. These results may be experimentally examined with a non-degenerate
unitary Bose gas, where the three-body recombination rate is substantially
reduced. In particular, the non-monotonic temperature dependence of
the two-body contact could be inferred from the momentum distribution
measurement.
\end{abstract}


\maketitle

\section{Introduction}

Understanding strongly interacting Bose gases in three dimensions
is a notoriously difficult quest \cite{Griffin2009,Griffin1996,Shi1998,Liu2004,Kita2009,Cooper2010,Cooper2011,Zhang2013,Yukalov2014}.
Theoretical studies of these systems have been hindered by the absence
of controllable theoretical approaches that can be used to describe
their properties within certain errors. Although a formal field-theoretical
description of weakly interacting Bose gases was developed more than
half a century ago by Lee, Huang and Yang \cite{Lee1957a,Lee1957b}
and later by Beliaev \cite{Beliaev1958} based on the ground-breaking
Bogoliubov theory \cite{Bogoliubov1947}. This theory is only applicable
in the limit of a small interaction parameter, the so-called gas parameter
$na_{s}^{3}\ll1$ - where $n$ is the density and $a_{s}>0$ is the
$s$-wave scattering length - as a result of the perturbative expansion.
When the gas parameter is extrapolated to infinity, each term appearing
in the perturbative field-theoretical description diverges. To the
best of our knowledge, a resummation of these divergent terms remains
unknown, even in an approximate manner.

Experimental studies, on the other hand, have been hampered by atom
losses from inelastic collisions. Unlike a strongly interacting Fermi
gas, where the atom loss rate due to three-body recombinations into
deeply bound diatomic molecules is greatly suppressed by the Pauli
exclusion principle \cite{Bloch2008}, at low temperatures an interacting
Bose gas has a three-body loss rate proportional to $a_{s}^{4}$ (i.e.,
the loss coefficient $\mathcal{L}_{3}\sim\hbar a_{s}^{4}/m$ \cite{Fedichev1996,Weber2003}),
which grows dramatically when $a_{s}$ is increased. Even in the absence
of inelastic collisions, for a strongly interacting Bose gas, the
possibility of recombination into deeply bound Efimov trimers \cite{Efimov1970}
indicates that the system can be at best metastable. 

Due to these realistic problems, experimental studies of a strongly
interacting atomic Bose gas near a broad Feshbach resonance have only
been carried out very recently \cite{Papp2008,Navon2011,Rem2013,Fletcher2013,Makotyn2014}.
The stability or lifetime of a unitary Bose gas with infinitely large
scattering length was investigated with $^{7}$Li \cite{Rem2013}
and $^{39}$K atoms \cite{Fletcher2013} in the non-degenerate regime.
It was found that there is a \emph{low-recombination} regime at high
temperatures and low densities, in which the loss coefficient saturates
at $L_{3}\sim\hbar\lambda_{dB}^{4}/m\propto1/T^{2}$, as predicted
\cite{DIncao2004}. Here, at high temperatures the thermal de Broglie
wavelength $\lambda_{dB}=[2\pi\hbar^{2}/(mk_{B}T)]^{1/2}$ replaces
the role of the $s$-wave scattering length $a_{s}$. The momentum
distribution of a quantum-degenerate unitary Bose gas was also measured
with $^{85}$Rb atoms \cite{Makotyn2014}. These rapid experimental
advances have trigged a number of interesting theoretical investigations
on the unitary Bose gas \cite{Cowell2002,Song2009,Lee2010,Diederix2011,Borzov2012,Li2012,Yin2013,Piatecki2014,Smith2014,Skyes2014,Jiang2014,Rossi2014,Laurent2014,Rossi2015,Ancilotto2015},
focusing particularly on the universal Bertsch parameter $\xi$, the
condensate fraction $n_{0}$ at zero temperature and quenching dynamics.
The predictions however are very different with each other, due to
the absence of an efficient theoretical framework to handle the intrinsic
strong correlations of a metastable unitary Bose gas.

In this work, we aim to develop a non-perturbative, controllable theory
of a strongly interacting Bose gas in its normal state, with an emphasis
on the high-temperature low-recombination regime in which our theoretical
predictions might be efficiently tested in future experiments. Our
description is built on an earlier innovative theoretical work by
Li and Ho \cite{Li2012}, who treated a repulsive Bose gas as a metastable
upper branch (defined later) of an interacting Bose gas across a broad
Feshbach resonance. By appropriately re-defining the upper branch
prescription \cite{He2015} and using a non-perturbative large-$N$
expansion approach to remove the unphysical non-linear effect in pair
fluctuations \cite{Nikolic2007,Veillette2007,Enss2012}, we overcome
the large mechanically unstable area encountered earlier at low temperatures
\cite{Li2012} and therefore make Li and Ho's idea practically useful
at arbitrary temperatures in the normal state and arbitrary interaction
strengths. Our improved theory is able to reproduce the path-integral
Monte-Carlo results at weak couplings \cite{Pilati2006} and the virial
expansion at high temperatures \cite{Liu2009,Liu2010a,Liu2010b,Castin2013,Liu2013}.
In the strongly interacting unitary limit, we calculate the equation
of state and Tan's two-body contact \cite{Tan2008} as a function
of temperature. An interesting non-monotonic temperature dependence
of the contact is predicted and is to be compared with future experimental
measurements of the momentum distribution.

The rest of the paper is organized as follows. In the next section,
we briefly introduce Li and Ho's idea of the upper branch Bose gas
and present the generalized Nozières-Schmitt-Rink (NSR) method. The
upper branch is then appropriately defined through an in-medium phase
shift. The large-$N$ expansion approach is adopted in order to overcome
the unphysical strong pair fluctuations at large interaction strengths.
In Sec. III, we first present the results for weakly interacting Bose
gases and compare them with the available path-integral Monte Carlo
simulations. We then discuss the equation of state and Tan's two-body
contact in the unitary limit. At sufficiently high temperatures, the
results are compared with the virial expansion predictions. Finally,
Sec. IV is devoted to the conclusions and outlooks.

\section{Generalized Nozières-Schmitt-Rink approach}

A three-dimensional (3D) interacting Bose gas can be described by
the imaginary-time action \cite{Popov1987}
\begin{equation}
\mathcal{S}=\int d\tau d\mathbf{x}\left[\bar{\psi}\left(\partial_{\tau}-\frac{\hbar^{2}}{2m}\nabla^{2}-\mu\right)\psi+\frac{U_{0}}{2}\bar{\psi}^{2}\psi^{2}\right],\label{eq:action}
\end{equation}
where $\bar{\psi}(x)$, $\psi(x)$ are $c$-number fields representing
the creation and annihilation operators of bosonic atoms of equal
mass $m$ at a space-time $x=(\mathbf{x},\tau)$. The imaginary time
$\tau$ runs from 0 to the inverse temperature $\beta=1/(k_{B}T)$
and $\mu$ is the chemical potential. The interatomic contact interaction
is parameterized by the bare strength $U_{0}<0$, which has to be
regularized by the two-particle $s$-wave scattering length $a_{s}$
via the relation,
\begin{equation}
\frac{1}{U_{0}}=\frac{m}{4\pi\hbar^{2}a_{s}}-\frac{1}{V}\sum_{\mathbf{k}}\frac{1}{2\epsilon_{\mathbf{k}}},
\end{equation}
where $V$ is the volume of the system and $\epsilon_{\mathbf{k}}\equiv\hbar^{2}\mathbf{k}^{2}/(2m)$
is the free-particle dispersion (i.e., kinetic energy). In experiment,
the scattering length $a_{s}$ can be conveniently tuned by a magnetic
field across a Feshbach resonance, to arbitrary values \cite{Bloch2008}.

It should be noted that in our model action, Eq. (\ref{eq:action}),
the contact interaction is always attractive ($U_{0}<0$), although
the scattering length can change sign across the Feshbach resonance.
This implies the pairing instability of two bosons and therefore the
ground state of the system would be a mixture of pairs and of the
remaining unpaired bosonic atoms \cite{Koetsier2009}, similar to
what happens for an interacting Fermi gas at the crossover from Bardeen-Cooper-Schrieffer
(BCS) superfluids to Bose-Einstein condensates (BEC) \cite{Eagles1969,Leggett1980,NSR1985,SadeMelo1993,Hu2006}.
In the normal state, such a mixture can be described by using the
seminal NSR approach \cite{NSR1985,SadeMelo1993,Hu2006}. 

Following the earlier work by Koetsier and co-workers \cite{Koetsier2009},
we introduce a pairing field 
\begin{equation}
\phi\left(\mathbf{x},\tau\right)=U_{0}\psi\left(\mathbf{x},\tau\right)\psi\left(\mathbf{x},\tau\right)
\end{equation}
and decouple the interatomic interaction via the standard Hubbard-Stratonovich
transformation, with which the atomic fields appear quadratically
and therefore can be formally integrated out. This leads to an effective
action for the pairing field and, at the level of Gaussian pair fluctuations,
results in the following grand thermodynamic potential:
\begin{eqnarray}
\Omega & = & \Omega_{0}+\delta\Omega,\label{eq:Omega}\\
\Omega_{0} & = & k_{B}T\sum_{\mathbf{k}}\ln\left(1-e^{-\beta\xi_{\mathbf{k}}}\right),\label{eq:Omega0}\\
\delta\Omega & = & k_{B}T\sum_{\mathbf{q},i\nu_{l}}\ln\left[-\Gamma^{-1}\left(\mathbf{q},i\nu_{l}\right)\right],\label{eq:OmegaParisFirst}
\end{eqnarray}
where $\xi_{\mathbf{k}}=\varepsilon_{{\bf k}}-\mu$. The last equation
is the contribution from pairs of bosons, which is characterized by
the two-particle vertex function (or the effective Green function
of pairs) $\Gamma\left(\mathbf{q},i\nu_{l}\right)$ with bosonic Matsubara
frequencies $\nu_{l}=2\pi lT$ ($l=0,\pm1,\pm2,\cdots$) \cite{Koetsier2009},
\begin{equation}
\Gamma^{-1}=\frac{m}{4\pi\hbar^{2}a_{s}}-\sum_{\mathbf{k}}\left[\frac{\gamma_{B}\left(q,k\right)}{i\nu_{l}-\xi_{\mathbf{q}/2+\mathbf{k}}-\xi_{\mathbf{q}/2-\mathbf{k}}}+\frac{1}{2\varepsilon_{\mathbf{k}}}\right].
\end{equation}
Here $n_{B}(x)=1/(e^{\beta x}-1)$ is the Bose-Einstein distribution
function and the factor $\gamma_{B}(\mathbf{q},\mathbf{k})\equiv1+n_{B}(\xi_{\mathbf{q}/2+\mathbf{k}})+n_{B}(\xi_{\mathbf{q}/2-\mathbf{k}})$
takes into account (in-medium) Bose enhancement of pair fluctuations.
By further converting the summation over Matsubara frequencies in
Eq. (\ref{eq:OmegaParisFirst}) into an integral over real frequency
and introducing an in-medium two-particle phase shift \cite{NSR1985,SadeMelo1993,Hu2006}
\begin{equation}
\delta\left(\mathbf{q},\omega\right)\equiv-{\rm Imln\left[-\Gamma^{-1}\left(\mathbf{q},\omega+i0^{+}\right)\right]},\label{eq:ps}
\end{equation}
the contribution to thermodynamic potential from the bosonic pairs
can be rewritten as 
\begin{equation}
\delta\Omega=-\frac{1}{\pi}\sum_{\mathbf{q}}\int_{-\infty}^{+\infty}d\omega\frac{1}{e^{\beta\omega}-1}\delta\left(\mathbf{q},\omega\right).\label{eq:OmegaPairs}
\end{equation}
To make the above integral meaningful, it is easy to see that, the
phase shift at zero frequency $\omega=0$ should vanish identically
for any momentum $\mathbf{q}$ because of the Bose-Einstein distribution
function. This is the so-called Thouless criterion, which is used
to determine the onset of pairing superfluidity. 

We note that, within the NSR approach, the only parameter in the imaginary-time
action - the chemical potential $\mu$ - is to be determined by using
the number equation,
\begin{equation}
n=-\frac{1}{V}\frac{\partial\left(\Omega_{0}+\delta\Omega\right)}{\partial\mu}\equiv n_{0}+\delta n,\label{eq:numberEQ}
\end{equation}
where $n$ is the number density of the system, consisting of both
the densities of atoms $n_{0}$ and of pairs $\delta n$.

For an attractive Bose gas near broad Feshbach resonances, Eq. (\ref{eq:Omega})
or Eq. (\ref{eq:numberEQ}) physically describes an ideal, non-interacting
mixture of bosonic atoms with destiny $n_{0}$ and pairs with density
$\delta n>0$. With increasing strength of attractive interactions,
the contribution from pairs, Eq. (\ref{eq:OmegaPairs}), becomes more
and more significant. As a result, the chemical potential decreases
to the half of the binding energy, $\mu\rightarrow-\hbar^{2}/(2ma_{s}^{2})$,
as required by the Thouless criterion $\delta\left(\mathbf{q},\omega\right)=0$
\cite{Koetsier2009}.

\begin{figure}
\begin{centering}
\includegraphics[clip,width=0.48\textwidth]{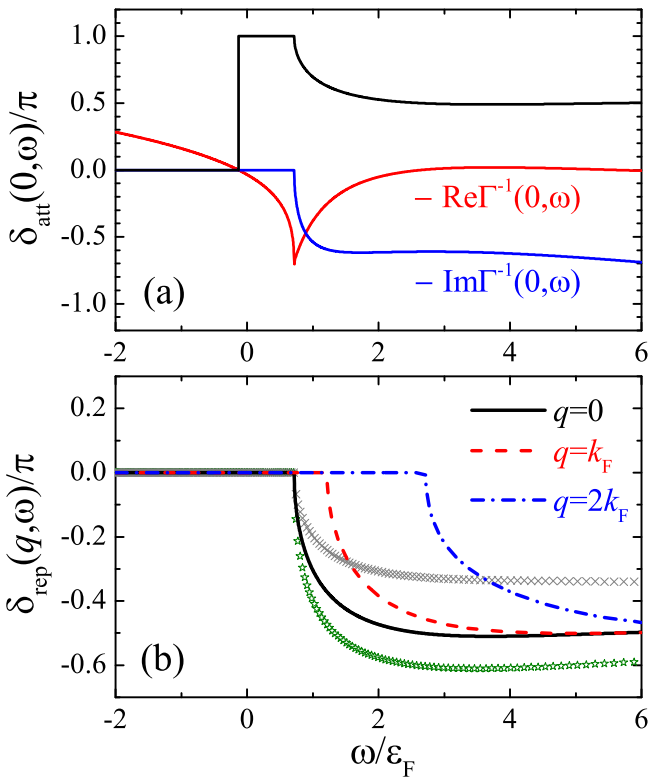} 
\par\end{centering}

\protect\caption{(Color online). (a) The in-medium phase shifts for an attractive Bose
gas $\delta_{\textrm{att}}(q=0,\omega)$ at the gas parameter $na_{s}^{3}=1$.
The real and imaginary parts of the negative inverse of the two-particle
vertex function, $-\Gamma^{-1}(q=0,\omega)$ are also shown. (b) The
corresponding in-medium phase shifts $\delta_{\textrm{rep}}(q,\omega)$
in the meta-stable upper branch at different momenta $q=0$ (black
solid line), $k_{F}$ (red dashed line) and $2k_{F}$ (blue dot-dashed
line). The phase shifts $\delta_{\textrm{rep}}(q=0,\omega)$ at different
gas parameters $na_{s}^{3}=0.01$ and $na_{s}^{3}=-1$ are shown by
gray cross and green stars, respectively. For all the plots, the temperature
is fixed at $T=2T_{c0}$, where $T_{c0}\simeq0.436T_{F}$ is the condensation
temperature of an ideal Bose gas. The chemical potential $\mu$ is
fixed to that of an ideal Bose gas at the same temperature, i.e.,
$\mu=\mu^{(0)}(2T_{c0})\simeq-0.358\varepsilon_{F}$.}

\label{fig1} 
\end{figure}

\subsection{In-medium phase shift for the ground state}

In Fig. \ref{fig1}(a), we show the typical behavior of the inverse
vertex function $\Gamma^{-1}$ and of the in-medium phase shift $\delta_{\textrm{att}}$
for an \emph{attractive} Bose gas with the gas parameter $na_{s}^{3}=1$
at $T=2T_{c0}$, where $T_{c0}\simeq0.436T_{F}$ is the condensation
temperature of an ideal Bose gas, measured in units of Fermi temperature
$T_{F}\equiv\hbar^{2}(3\pi^{2}n)^{2/3}/(2mk_{B})\equiv\varepsilon_{F}/k_{B}$.
The phase shift jumps from zero to $\pi$ at the threshold frequency
$\omega_{b}(\mathbf{q})$, which signals the existence of bound states.
Upon increasing the frequency beyond the scattering threshold, 
\begin{equation}
\omega_{s}\left(\mathbf{q}\right)=\frac{\hbar^{2}\mathbf{q}^{2}}{4m}-2\mu,
\end{equation}
where the imaginary part of the vertex function becomes nonzero, the
phase shift decreases towards $\pi/2$ as $\omega\rightarrow+\infty$.
Therefore, there are two contributions to the phase shift, originating
from the bound states (at $\omega_{b}(\mathbf{q})\leq\omega<\omega_{s}(\mathbf{q})$)
and from the scattering states (i.e., $\omega\geq\omega_{s}(\mathbf{q})$),
respectively. It is clear from Fig. \ref{fig1}(a) that the phase
shift, as an illustrated example, does not satisfy the constraint
$\delta\left(\mathbf{q},\omega=0\right)=0$. This is because we have
used an artificially large chemical potential, larger than the actual
chemical potential, which has to be solved self-consistently by using
the number equation (\ref{eq:numberEQ}) for an attractive Bose gas.

\subsection{In-medium phase shift for the upper branch}

It is interesting that although we are dealing with an attractive
Bose gas, we may also obtain useful information about a repulsively
interacting Bose gas, by treating it as a metastable upper branch
of the attractive system. This idea may be understood from the fact
that there is an ambiguity in calculating the in-medium phase shift
Eq. (\ref{eq:ps}), as it involves a multi-valued $\ln(x)$ function.
By appropriately choosing different branch cuts, one thus may access
excited many-body states, in addition to the ground state of the system.

To the best of our knowledge, the proper choice of in-medium phase
shift was first emphasized by Engelbrecht and Randeria in the study
of a weakly interacting repulsive Fermi gas in two dimensions in 1992
\cite{Engelbrecht1992}. However, at that time, the connection between
attractive and repulsive systems was not realized and the concept
of the upper branch was not established. The meaning of the upper
branch was only clarified in 2011 by Shenoy and Ho, who claimed that
by excluding the contribution from the paired molecular states in
calculating the thermodynamics of the system, one could access the
upper branch of an attractive Fermi gas \cite{Shenoy2011}. This excluded
molecular pole approximation (EMPA) immediately implies that for the
upper branch, the lower boundary of the frequency integral in Eq.
(\ref{eq:OmegaPairs}) should be modified to $\omega_{s}(\mathbf{q})$,
leading to 
\begin{equation}
\delta\Omega=-\frac{1}{\pi}\sum_{\mathbf{q}}\int_{\omega_{s}\left(\mathbf{q}\right)}^{+\infty}d\omega\frac{1}{e^{\beta\omega}-1}\delta_{\textrm{rep}}\left(\mathbf{q},\omega\right).\label{eq:OmegaPairsRep}
\end{equation}
This expression was later applied by Li and Ho to a strongly interacting
Bose gas \cite{Li2012}. However, despite the clarification of the
concept of the upper branch, in those two studies (i.e., Refs. \cite{Li2012}
and \cite{Shenoy2011}), the ambiguity in the calculation of the phase
shift $\delta_{\textrm{rep}}(\mathbf{q},\omega)$ was not carefully
treated. The phase shift of the upper branch was directly calculated
by using 
\begin{equation}
\delta_{\textrm{rep}}^{\textrm{HO}}\left(\mathbf{q},\omega\geq\omega_{s}\left(\mathbf{q}\right)\right)=-\arctan\left[\frac{\textrm{Im\ensuremath{\Gamma^{-1}\left(\mathbf{q},\omega\right)}}}{\textrm{Re\ensuremath{\Gamma^{-1}\left(\mathbf{q},\omega\right)}}}\right]\label{eq:psRepHO}
\end{equation}
without the explanation for the branch cut. Here, the function $\arctan(x)$
is the usual inverse tangent function that takes values in the first
and fourth quadrant $(-\pi/2,+\pi/2)$ \cite{note1} and we have used
the superscript ``HO'' to indicate the prescription given by Ho
and co-workers.

It turns out that a more appropriate phase shift for the upper branch
can be physically defined by the prescription
\begin{equation}
\delta_{\textrm{rep}}\left(\mathbf{q},\omega\right)=\left[\delta_{\textrm{att}}\left(\mathbf{q},\omega\right)-\pi\right]\Theta\left[\omega-\omega_{s}\left(\mathbf{q}\right)\right],\label{eq:psRep}
\end{equation}
which can be shown from the viewpoint of the virial expansion \cite{He2015}.
The $\pi$-shift in the above equation can be easily understood from
the standard scattering theory: when a two-body bound state emerges,
the two-particle phase shift associated with the density of states
should increase by $\pi$. The prescription Eq. (\ref{eq:psRep})
is therefore simply the many-body generalization of the two-particle
phase shift in the absence of bound states. It should be viewed as
a physical realization of the EMPA approximation proposed by Ho and
co-workers.

For a weakly interacting Bose gas (i.e., $a_{s}\rightarrow0^{+}$),
The two prescriptions for the upper branch phase shift, shown in Eqs.
(\ref{eq:psRepHO}) and (\ref{eq:psRep}), agree with each other,
as a result of the large value of $\textrm{Re\ensuremath{\Gamma^{-1}\left(\mathbf{q},\omega\right)}}$.
Towards the strongly interacting limit $a_{s}\rightarrow+\infty$,
however, the two prescriptions differ significantly. In particular,
on the BCS side with a negative scattering length, the phase shift
$\delta_{\textrm{rep}}^{\textrm{HO}}(\mathbf{q},\omega)$ coincides
with the phase shift of the ground state branch, $\delta_{\textrm{att}}(\mathbf{q},\omega)$.
As a result, by changing the scattering length and crossing the Feshbach
resonance from below, there is a sudden branch switch from the upper
branch to the ground state branch \cite{Li2012}. This branch-switching
effect and the related violation of exact Tan's relations \cite{Li2012}
are absent when the more physical prescription Eq. (\ref{eq:psRep})
is used. 

In Fig. \ref{fig1}(b), we show the in-medium phase shift for the
upper branch, obtained by performing Eq. (\ref{eq:psRep}) for the
attractive phase shift shown in Fig. \ref{fig1}(a). The Thouless
criterion $\delta\left(\mathbf{q},\omega=0\right)=0$ is now strictly
satisfied. Moreover, for a positive frequency the phase shift becomes
negative. This leads to a positive fluctuation thermodynamic potential
$\delta\Omega>0$ and a negative pair density $\delta n<0$. As analyzed
by Engelbrecht and Randeria \cite{Engelbrecht1992}, the fact $\delta\Omega>0$
implies that the ground-state energy increases due to the interactions,
as it should be for a repulsive system. A negative pairing density
is also consistent with a repulsive interaction, which, for a specific
atom, will expel other atoms away from its position, and therefore
make the effective number density around its position smaller.

In Fig. \ref{fig1}(b), we also report the upper branch phase shift
at the gas parameter $na_{s}^{3}=0.01$ (grey crosses) and $na_{s}^{3}=-1$
(green stars). It is worth noting that the negative value of the gas
parameter (i.e., on the BCS side above the Feshbach resonance) actually
means stronger repulsions between atoms, as indicated by the large
absolute value of the phase shift. In contrast, for a positive gas
parameter, the interaction effect becomes weaker with decreasing the
gas parameter.

\subsection{Large-$N$ expansion}

The generalized NSR approach was used earlier by Li and Ho to investigate
a strongly interacting Bose gas near unitarity \cite{Li2012}. A large
mechanically unstable area was found when the temperature of the system
is below $T<5T_{c0}\sim2T_{F}$, which renders the approach useful
only at extremely high temperatures. Here, we show that the mechanical
instability is artificial and caused by the inappropriate treatment
for the strong pair/density fluctuations in the NSR approach. It can
be cured by the so-called large-$N$ expansion technique \cite{Nikolic2007,Veillette2007,Enss2012}.

In the large-$N$ expansion, we assign an additional flavor degree
of freedom to bosonic atoms ($i,j=1,\cdots,N$) and thereby extend
the model action to,
\begin{eqnarray}
\mathcal{\tilde{S}} & = & \int d\tau d\mathbf{x}\left[\sum_{i=1}^{N}\bar{\psi_{i}}\left(\partial_{\tau}-\frac{\hbar^{2}}{2m}\nabla^{2}-\mu\right)\psi_{i}\right.\nonumber \\
 &  & \left.+\frac{U_{0}}{2N}\sum_{i,j=1}^{N}\bar{\psi}_{i}^{2}\left(\mathbf{x},\tau\right)\psi_{j}^{2}\left(\mathbf{x},\tau\right)\right].\label{eq:actionLargeN}
\end{eqnarray}
By introducing a pairing field 
\begin{equation}
\tilde{\phi}\left(\mathbf{x},\tau\right)=\frac{U_{0}}{N}\sum_{i=1}^{N}\psi_{i}\left(\mathbf{x},\tau\right)\psi_{i}\left(\mathbf{x},\tau\right)
\end{equation}
and again decoupling the interatomic interaction via the standard
Hubbard-Stratonovich transformation, we integrate out the atomic fields
and obtain the grand thermodynamic potential per flavor 
\begin{equation}
\frac{\tilde{\Omega}}{N}=\Omega_{0}+\frac{1}{N}\delta\Omega+O(\frac{1}{N^{2}}),\label{eq:OmegaLargeN}
\end{equation}
up to the first non-trivial order of $O(1/N)$ \cite{Nikolic2007,Veillette2007,Enss2012}.
Here, for the metastable upper branch, $\Omega_{0}$ and $\delta\Omega$
are given by Eqs. (\ref{eq:Omega0}) and (\ref{eq:OmegaPairsRep}),
respectively. 

It is clear that in the large-$N$ expansion we have introduced an
artificial small parameter $1/N$, which can be used to control the
accuracy of the theory of strongly interacting Bose gases. The NSR
approach, which is based on the summation of infinite ladder diagrams
\cite{NSR1985,Hu2006}, should be understood as an approximate theory
obtained by directly setting $N=1$. However, such a procedure cannot
be justified \emph{a priori} in the strongly interacting regime, as
the controllable parameter $1/N$ is already at the order of unity.
Indeed, the appearance of the large mechanically unstable regime at
low temperatures, mentioned at the beginning of this subsection, is
precisely an indication of the breakdown of the procedure of directly
setting $N=1$. A more reasonable treatment is to first solve the
thermodynamics of a $N$-flavor system with $N\gg1$ and then \emph{linearly}
extrapolate all the desired physical quantities - as a function of
$1/N$ - to the limit of $N=1$. This large-$N$ expansion idea has
been successfully applied to a strongly interacting two-component
Fermi gas in the unitary limit \cite{Nikolic2007,Veillette2007}.
The equation of state and the Tan contact near the quantum critical
point $\mu=0$ was then accurately predicted \cite{Enss2012}. In
this work, we anticipate that the same large-$N$ expansion technique
could also lead to very useful information for a unitary Bose gas
in the quantum critical region.

\begin{figure}[t]
\begin{centering}
\includegraphics[clip,width=0.48\textwidth]{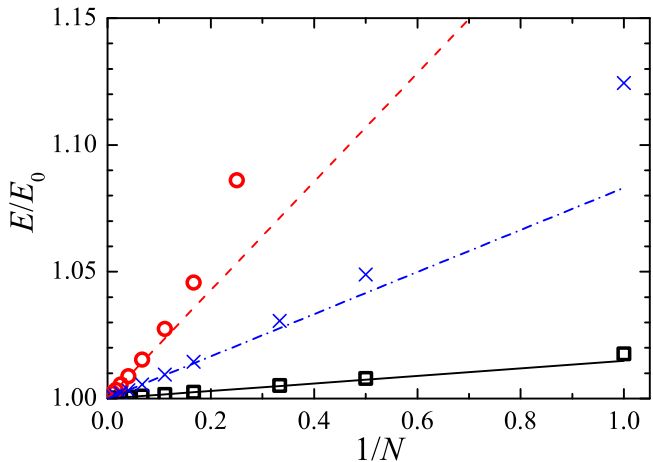} 
\par\end{centering}

\protect\caption{(Color online) Energy as a function of the artificial controlling
parameter $1/N$ at $T=2T_{c0}$ (black squares for $na_{s}^{3}=10^{-6}$
and red circles for $na_{s}^{3}=1$) and $T=10T_{c0}$ (blue crosses
with $na_{s}^{3}=1$). The energy is measured in units of the energy
$E_{0}$ of a non-interacting Bose gas at the same temperature. The
lines are linear fits to the small $1/N$ data. Note that, at low
temperatures and strong interactions (i.e., red circles), we are not
able to find solutions for $N=1$ because of strong correlations.}

\label{fig2} 
\end{figure}

In Fig. \ref{fig2}, we show the $1/N$-dependence of the total energy
of an interacting Bose gas at different gas parameters and temperatures,
obtained by solving the coupled equations Eqs. (\ref{eq:Omega0}),
(\ref{eq:OmegaPairsRep}) and (\ref{eq:OmegaLargeN}), and subject
to the number equation $\tilde{n}/N\equiv n=n_{0}+\delta n/N$ for
the number density per flavor $\tilde{n}/N$. At weak interactions
(black squares) or high temperatures (blue crosses), roughly the energy
changes linearly as a function of $1/N$. The linear extrapolation
approximation used in the large-$N$ expansion therefore does not
make significant difference. However, for a strongly interacting Bose
gas at relatively low temperatures (red circles), the dependence is
highly non-linear. In particular, we are not able to find physical
solutions when the number of flavors $N\leq2$. Therefore, it becomes
crucial to keep only the linear term in the $1/N$ expansion, which
provides the first non-trivial and non-pertrubative knowledge about
a strongly correlated many-body state.

\section{Results and discussions}

\begin{figure}
\begin{centering}
\includegraphics[clip,width=0.48\textwidth]{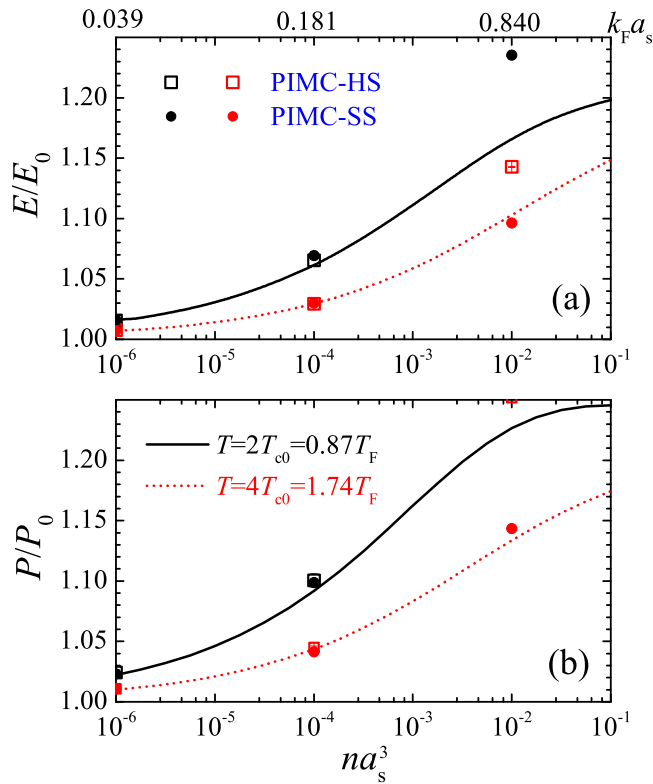} 
\par\end{centering}

\protect\caption{(Color online) The energy (a) and pressure (b) as a function of the
gas parameter $na_{s}^{3}$ at $T=2T_{c0}$ (black solid lines) and
$T=4T_{c0}$ (red dashed lines), normalized respectively by their
corresponding results of an ideal, non-interacting Bose gas at the
same temperature. The results from a path-integral Monte-Carlo calculations
are also shown \cite{Pilati2006}, with squares for hard-sphere potential
and circles for soft-sphere potential.}

\label{fig3} 
\end{figure}

In this section, we present our large-$N$ results, calculated by
the linear extrapolation towards the limit $1/N=1$. In practice,
we solve the generalized NSR approach with $N=50-100$ for the chemical
potential $\mu(N)$ and the energy $E(N)$, and then expand them around
the corresponding non-interacting values $\mu_{0}$ and $E_{0}$,
\begin{eqnarray}
\mu\left(N\right) & = & \mu_{0}+\delta\mu/N+O\left(1/N^{2}\right),\\
E\left(N\right) & = & E_{0}+\delta E/N+O\left(1/N^{2}\right),
\end{eqnarray}
to extract the corrections $\delta\mu$ and $\delta E$. This leads
to the large-$N$ expansion results $\mu=\mu_{0}+\delta\mu$ and $E=E_{0}+\delta E$.

\subsection{Crossover to strong repulsions}

In Figs. \ref{fig3}(a) and \ref{fig3}(b), we present respectively
the energy and pressure of an interacting Bose gas at two temperatures
$T=2T_{c0}$ (black solid lines) and $T=4T_{c0}$ (red dashed lines),
as a function of the gas parameter $na_{s}^{3}$, or $k_{F}a_{s}$
if we convert the number density $n$ to a Fermi wavevector $k_{F}=(3\pi^{2}n)^{1/3}$.
The large-$N$ expansion results are compared with available path-integral
Monte Carlo calculations for a hard-sphere (squares) and soft-sphere
potential (circles) \cite{Pilati2006}. For weak interactions (i.e.,
$na_{s}^{3}=10^{-6}$ and $10^{-4}$ or $k_{F}a_{s}<0.2$), our predictions
agree well with the \emph{ab-initio} simulations. For strong interactions
with strength $k_{F}a_{s}\sim0.8$, there is a significant difference.
This is due to the effect of non-negligible effective range of interactions
$r_{0}$ used in the Monte Carlo simulations (i.e., $\left|k_{F}r_{0}\right|\sim1$),
which leads to a sizable correction to the energy and pressure. In
our calculations with a contact interaction, the range of interactions
is strictly zero.

\begin{figure}
\begin{centering}
\includegraphics[clip,width=0.48\textwidth]{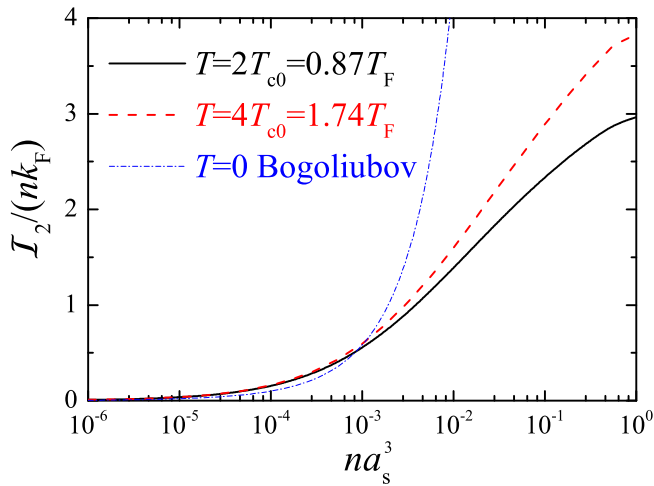} 
\par\end{centering}

\protect\caption{(Color online) Two-body contact $\mathcal{I}_{2}$ as a function of
the gas parameter $na_{s}^{3}$ at $T=2T_{c0}$ (black solid line)
and $T=4T_{c0}$ (red dashed line). The result from the zero-temperature
Bogoliubov theory is shown by the blue dot-dashed line.}

\label{fig4} 
\end{figure}

\begin{figure}
\begin{centering}
\includegraphics[clip,width=0.48\textwidth]{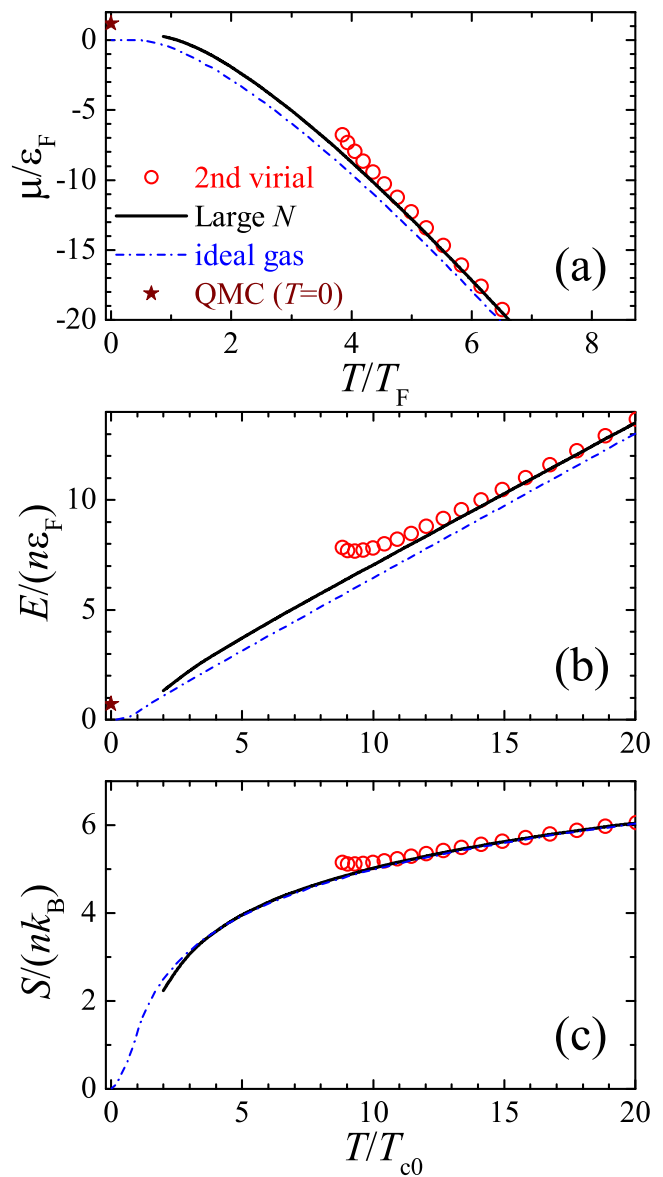} 
\par\end{centering}

\protect\caption{(Color online) Temperature dependence of the chemical potential (a),
energy (b) and entropy (c) of a unitary Bose gas. For comparison,
we show the second-order virial expansion predictions by red empty
circles and the ideal gas results by dot-dashed lines. The latest
QMC results at zero temperature are also plotted by using stars in
brown \cite{Rossi2014}. }

\label{fig5} 
\end{figure}

In Fig. \ref{fig4}, we show the evolution of the corresponding Tan's
contact with increasing gas parameter. Tan's contact measures the
density of pairs at short distance and determines the exact large-momentum
or high-frequency behavior of various physical observables \cite{Tan2008}.
It therefore serves as an important quantity to characterize a strongly
interacting many-body system. In particular, experimentally it can
be measured from the momentum distribution \cite{Makotyn2014,Wild2012},
which takes a $k^{-4}$ tail in the short-wavelength limit, i.e.,
$n(\mathbf{k})\rightarrow\mathscr{\mathcal{I}}/\mathbf{k}^{4}$. At
finite temperatures, the contact can be conveniently calculated by
using the adiabatic relation \cite{Tan2008,Hu2011NJP}:
\begin{equation}
\mathcal{I}_{2}=-\frac{4\pi m}{\hbar^{2}}\left[\frac{\partial\Omega}{\partial a_{s}^{-1}}\right]_{T,\mu}.\label{eq:TanRelation}
\end{equation}
We have used the subscript ``$2$'' to emphasize the fact that in
our calculations we do not consider the three-body Efimov physics
and the associated inelastic collisions. These effects are instead
captured by a three-body contact $\mathcal{I}_{3}$, which can be
defined through an adiabatic relation for a three-body parameter.
We refer to Ref. \cite{Smith2014} for more detailed discussions. 

The two-body contact is an increasing function of the interaction
strength. At small gas parameters, our results are in good agreement
with the weak-coupling predictions of a zero-temperature Bogoliubov
theory (thin blue dot-dashed line) \cite{Schakel2010}, 
\begin{equation}
\mathcal{I}_{\textrm{bog}}=\left(4\pi na_{s}\right)^{2}\left[1+\frac{64}{3\sqrt{\pi}}\sqrt{na_{s}^{3}}\right].
\end{equation}
The slight increase in our large-$N$ expansion results is due to
the finite temperature effect. At large gas parameters $na_{s}^{3}$,
the contact tends to saturate to a universal value that depends only
on the temperature, as it should be.

\subsection{Unitary Bose gases}

We are now in position to discuss the universal thermodynamics of
a unitary Bose gas. In Fig. \ref{fig5}, we present the chemical potential,
energy and entropy, as a function of temperature. For comparison,
we also plot in dot-dashed lines the temperature dependence of an
ideal, non-interacting Bose gas. For the chemical potential and energy,
our results lie systematically above the non-interacting results,
clearly indicating the consequence of strong repulsions. They tends
to converge to the zero temperature quantum Monte Carlo predictions
(brown stars) with decreasing the temperature \cite{Rossi2014}. In
contrast, the entropy seems to be less affected by strong repulsions.
The insensitivity of entropy on the interatomic interactions was also
previously found for a unitary Fermi gas \cite{Hu2008,Hu2010,Hu2011PRA}.

At high temperatures with a small fugacity $z=e^{\beta\mu}\ll1$,
we may use the virial expansion theory to study the universal thermodynamics
\cite{Liu2013}. For a unitary Bose gas, the virial expansion of the
grand thermodynamic potential takes the form,
\begin{eqnarray}
\Omega & = & \Omega_{0}-\frac{k_{B}T}{\lambda_{dB}^{3}}\left(z^{2}\Delta b_{2}+z^{3}\Delta b_{3}+\cdots\right),\label{eq:OmegaVE}
\end{eqnarray}
where $\lambda_{dB}\equiv[2\pi\hbar^{2}/(mk_{B}T)]^{1/2}$ is the
thermal de-Broglie wavelength, and $\Delta b_{2}=-\sqrt{2}$ is the
second-order virial coefficient for strong repulsions \cite{Castin2013,note2},
which can be easily calculated by using Beth-Uhlenbeck formalism \cite{Beth1937}.
Up to the second order, we can solve Eq. (\ref{eq:OmegaVE}) together
with the number equation Eq. (\ref{eq:numberEQ}). For the fugacity,
we find that
\begin{eqnarray}
z & \simeq & \sqrt{\frac{16}{9\pi}}\left(\frac{T_{F}}{T}\right)^{3/2}\label{eq:zVE}
\end{eqnarray}
at $T\gg T_{F}$. The virial predictions for the equation of state
are shown in Fig. \ref{fig5} by red circles and agree well with the
large-$N$ expansion results at high temperatures.

\begin{figure}
\begin{centering}
\includegraphics[clip,width=0.48\textwidth]{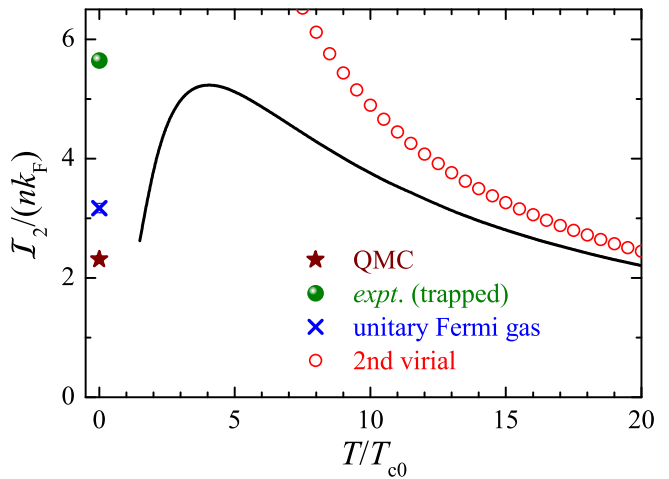} 
\par\end{centering}

\protect\caption{(Color online) Temperature dependence of the two-body contact $\mathcal{I}_{2}$
in the unitary limit. The prediction from a second-order virial expansion,
Eq. (\ref{eq:contactVE}), is shown by red empty circles. At zero
temperature, the brown star and green solid circle refer to the latest
QMC result \cite{Rossi2014} and the number extracted from the recent
measurement at JILA for a trapped unitary Bose gas \cite{Makotyn2014,Smith2014}.
The blue cross indicates the zero-temperature contact of a unitary
Fermi gas, measured very precisely by using Bragg spectroscopy \cite{Hoinka2013}.}

\label{fig6} 
\end{figure}
In the unitary limit, we may calculate the universal contact by using
the adiabatic relation, Eq. (\ref{eq:TanRelation}), shown in Fig.
\ref{fig6}. With increasing temperature, the contact initially increases
and then decreases, giving rise to a peak structure at the temperature
$T\sim4T_{c0}$. The decrease of the contact at high temperatures
can be well understood by using the virial expansion theory for the
contact. As a direct consequence of the adiabatic relation, we have
the expansion,
\begin{eqnarray}
\mathcal{I}_{2} & = & \frac{8\pi m}{\hbar^{2}}\frac{k_{B}T}{\lambda_{dB}^{2}}\left(z^{2}c_{2}+z^{3}c_{3}+\cdots\right),
\end{eqnarray}
where $c_{n}=\lambda_{dB}^{-1}(\partial\Delta b_{n}/\partial a_{s}^{-1})$
is the so-called contact coefficient \cite{Hu2011NJP}. For a unitary
Bose gas, by using Beth-Uhlenbeck formalism it is easy to show that
$c_{2}=2/\pi$. Using Eq. (\ref{eq:zVE}), we then find
\begin{eqnarray}
\frac{I_{2}}{nk_{F}} & \simeq & \frac{64}{3}\left(\frac{T_{F}}{T}\right).\label{eq:contactVE}
\end{eqnarray}
Thus, at high temperatures the contact decreases as $T^{-1}$. There
is no apparent physical explanation for the increase of the contact
at low temperatures. However, we notice that with decreasing temperature
toward zero temperature, our large-$N$ expansion result seems to
be consistent with the zero-temperature value predicted by the latest
quantum Monte Carlo simulation \cite{Rossi2014}. 

For comparison, we show the experimental data of the contact \cite{Makotyn2014},
analyzed by Smith and co-workers (green solid circle) \cite{Smith2014}
. The significant discrepancy between experiment and our large-$N$
theory should be largely due to the unknown temperature in the experiment,
as the Bose cloud could be significantly heated by atom losses \cite{Makotyn2014}.
We also show the zero-temperature contact of a unitary Fermi gas (blue
cross), which has been both calculated and measured very accurately
\cite{Hoinka2013}. It is interesting that both the unitary Bose and
Fermi gases have similar contact at zero temperature, indicating that
a 3D Bose gas may also have the tendency of being fermionized at strong
repulsions, analogous to a Bose gas in one dimension.

\section{Conclusions}

In summary, based on the upper branch idea and large-$N$ expansion
technique, we have developed a unified theory for a normal, strongly
interacting Bose gas. The theory reproduces the path-integral Monte
Carlo simulation in the weak-coupling limit \cite{Rossi2014}. While
at high temperatures, it nicely recovers the known results from a
quantum virial expansion calculation \cite{Castin2013,Liu2013}. Thus,
we anticipate that the universal thermodynamics predicted by our theory
could be qualitatively reliable. A useful check may be provided by
experimentally measuring the finite-temperature contact of a unitary
Bose gas through momentum distribution or momentum-resolved radio-frequency
spectroscopy \cite{Makotyn2014}.

Our results complement the earlier studies of a condensed strongly
interacting Bose gas. It is worth nothing that our theoretical framework
can naturally be extended to include the condensation (i.e., $T<T_{c0}$)
by using a generalized Nozières-Schmitt-Rink approach below the superfluid
transition temperature \cite{Hu2006}. This extension will be addressed
in a future investigation.
\begin{acknowledgments}
X.-J.L. and H.H. acknowledge the support from the ARC Discovery Projects
(Grant Nos. FT140100003, DP140100637, FT130100815 and DP140103231)
and the National Key Basic Research Special Foundation of China (NKBRSFC-China)
(Grant No. 2011CB921502). L.H. was supported by the US Department
of Energy Nuclear Physics Office (Contract No. DOE-AC02-05CH11231).\end{acknowledgments}


\begin{thebibliography}{10}
\bibitem{Griffin2009}A. Griffin, T. Nikuni, and E. Zaremba, \textit{Bose-Condensed
Gases at Finite Temperatures} (Cambridge University Press, Cambridge,
UK, 2009).

\bibitem{Griffin1996}A. Griffin, Phys. Rev. B \textbf{53}, 9341 (1996).

\bibitem{Shi1998}H. Shi and A. Griffin, Phys. Rep. \textbf{304},
1 (1998).

\bibitem{Liu2004}X.-J. Liu, H. Hu, A. Minguzzi, and M. P. Tosi, Phys.
Rev. A \textbf{69}, 043605 (2004).

\bibitem{Kita2009}T. Kita, Phys. Rev. B \textbf{80}, 214502 (2009).

\bibitem{Cooper2010}F. Cooper, C.-C. Chien, B. Mihaila, J. F. Dawson,
and E. Timmermans, Phys. Rev. Lett. \textbf{105}, 240402 (2010).

\bibitem{Cooper2011}F. Cooper, B. Mihaila, J. F. Dawson, C.-C. Chien,
and E. Timmermans, Phys. Rev. A \textbf{83}, 053622 (2011).

\bibitem{Zhang2013}Y.-H. Zhang and D. Li, Phys. Rev. A \textbf{88},
053604 (2013).

\bibitem{Yukalov2014}V. I. Yukalov and E. P. Yukalova, Phys. Rev.
A \textbf{90}, 013627 (2014).

\bibitem{Lee1957a}T. D. Lee and C. N. Yang, Phys. Rev. \textbf{105},
1119 (1957).

\bibitem{Lee1957b}T. D. Lee, K. Huang, and C. N. Yang, Phys. Rev.
\textbf{106}, 1135 (1957).

\bibitem{Beliaev1958}S. T. Beliaev, Sov. Phys. JETP \textbf{7}, 289
(1958); \textbf{7}, 299 (1958).

\bibitem{Bogoliubov1947}N. N. Bogoliubov, J. Phys. (USSR) \textbf{11},
23 (1947).

\bibitem{Bloch2008}I. Bloch, J. Dalibard, and W. Zwerger, Rev. Mod.
Phys. \textbf{80}, 885 (2008).

\bibitem{Fedichev1996}P. O. Fedichev, M. W. Reynolds, and G. V. Shlyapnikov,
Phys. Rev. Lett. \textbf{77}, 2921 (1996).

\bibitem{Weber2003}T. Weber, J. Herbig, M. Mark, H.-C. Nägerl, and
R. Grimm, Phys. Rev. Lett. \textbf{91}, 123201 (2003).

\bibitem{Efimov1970}V. Efimov, Phys. Lett. \textbf{33B}, 563 (1970).

\bibitem{Papp2008}S. B. Papp, J. M. Pino, R. J. Wild, S. Ronen, C.
E. Wieman, D. S. Jin, and E. A. Cornell, Phys. Rev. Lett. \textbf{101},
135301 (2008).

\bibitem{Navon2011}N. Navon, S. Piatecki, K. Günter, B. S. Rem, T.
C. Nguyen, F. Chevy, W. Krauth, and C. Salomon, Phys. Rev. Lett. \textbf{107},
135301 (2011).

\bibitem{Rem2013}B. S. Rem, A. T. Grier, I. Ferrier-Barbut, U. Eismann,
T. Langen, N. Navon, L. Khaykovich, F. Werner, D. S. Petrov, F. Chevy,
and C. Salomon, Phys. Rev. Lett. \textbf{110}, 163202 (2013).

\bibitem{Fletcher2013}R. J. Fletcher, A. L. Gaunt, N. Navon, R. P.
Smith, and Z. Hadzibabic, Phys. Rev. Lett. \textbf{111}, 125303 (2013).

\bibitem{Makotyn2014}P. Makotyn, C. E. Klauss, D. L. Goldberger,
E. A. Cornell, and D. S. Jin, Nature Phys. \textbf{10}, 116 (2014).

\bibitem{DIncao2004}J. P. D\textquoteright Incao, H. Suno, and B.
D. Esry, Phys. Rev. Lett. \textbf{93}, 123201 (2004).

\bibitem{Cowell2002}S. Cowell. H. Heiselberg, I. E. Mazets, J. Morales,
V. R. Pandharipande, and C. J. Pethick, Phys. Rev. Lett. \textbf{88},
210403 (2002).

\bibitem{Song2009}J.-L. Song and F. Zhou, Phys. Rev. Lett. \textbf{103},
025302 (2009).

\bibitem{Lee2010}Y.-L. Lee and Y.-W. Lee, Phys. Rev. A \textbf{81},
063613 (2010).

\bibitem{Diederix2011}J. M. Diederix, T. C. F. van Heijst, and H.
T. C. Stoof, Phys. Rev. A \textbf{84}, 033618 (2011).

\bibitem{Borzov2012}D. Borzov, M. S. Mashayekhi, S. Zhang, J.-L.
Song, and F. Zhou, Phys. Rev. A \textbf{85}, 023620 (2012).

\bibitem{Li2012}W. Li and T.-L. Ho, Phys. Rev. Lett. \textbf{108},
195301 (2012).

\bibitem{Yin2013}X. Yin and L. Radzihovsky, Phys. Rev. A \textbf{88},
063611 (2013).

\bibitem{Piatecki2014}S. Piatecki and W. Krauth, Nat. Commun. \textbf{5},
3503 (2014).

\bibitem{Smith2014}D. H. Smith, E. Braaten, D. Kang, and L. Platter,
Phys. Rev. Lett. \textbf{112}, 110402 (2014).

\bibitem{Skyes2014}A. G. Skyes, J. P. Corson, J. P. D'Incao, A. P.
Koller, C. H. Greene, A. M. Rey, K. R. A. Hazzard, and J. L. Bohn,
Phys. Rev. A \textbf{89}, 021601(R) (2014).

\bibitem{Jiang2014}S.-J. Jiang, W.-M. Liu, G. W. Semenoff, and F.
Zhou, Phys. Rev. A \textbf{89}, 033614 (2014).

\bibitem{Rossi2014}M. Rossi, L. Salasnich, F. Ancilotto, and F. Toigo,
Phys. Rev. A \textbf{89}, 041602(R) (2014).

\bibitem{Laurent2014}S. Laurent, X. Leyronas, and F. Chevy, Phys.
Rev. Lett. \textbf{113}, 220601 (2014).

\bibitem{Rossi2015}M. Rossi, F. Ancilotto, L. Salasnich, and F. Toigo,
arXiv:1408.3945 (2014).

\bibitem{Ancilotto2015}F. Ancilotto, M. Rossi, L. Salasnich, and
F. Toigo, arXiv:1501.0549 (2015).

\bibitem{He2015}L. He, X.-J. Liu, X.-G. Huang, and H. Hu, arXiv:1412.2412
(2014).

\bibitem{Nikolic2007}P. Nikoli\'{c} and S. Sachdev, Phys. Rev. A
\textbf{75}, 033608 (2007).

\bibitem{Veillette2007}M. Y. Veillette, D. E. Sheehy, and L. Radzihovsky,
Phys. Rev. A \textbf{75}, 043614 (2007).

\bibitem{Enss2012}T. Enss, Phys. Rev. A \textbf{86}, 013616 (2012).

\bibitem{Pilati2006}S. Pilati, K. Sakkos, J. Boronat, J. Casulleras,
and S. Giorgini, Phys. Rev. A \textbf{74}, 043621 (2006).

\bibitem{Liu2009}X.-J. Liu, H. Hu, and P. D. Drummond, Phys. Rev.
Lett. \textbf{102}, 160401 (2009).

\bibitem{Liu2010a}X.-J. Liu, H. Hu, and P. D. Drummond, Phys. Rev.
A \textbf{82}, 023619 (2010).

\bibitem{Liu2010b}X.-J. Liu, H. Hu, and P. D. Drummond, Phys. Rev.
B \textbf{82}, 054524 (2010).

\bibitem{Castin2013}Y. Castin and F. Werner, Canadian Journal of
Physics \textbf{91}, 382 (2013).

\bibitem{Liu2013}X.-J. Liu, Phys. Rep. \textbf{524}, 37 (2013).

\bibitem{Tan2008}S. Tan, Ann. Phys. (NY) \textbf{323}, 2952 (2008);
\textbf{323}, 2971 (2008).

\bibitem{Popov1987}V. N. Popov, \textit{Functional Integrals and
Collective Excitations} (Cambridge University Press, Cambridge, UK,
1987).

\bibitem{Koetsier2009}A. Koetsier, P. Massignan, R. A. Duine, and
H. T. C. Stoof, Phys. Rev. A \textbf{79}, 063609 (2009).

\bibitem{Eagles1969}D. M. Eagles, Phys. Rev. \textbf{186}, 456 (1969).

\bibitem{Leggett1980}A. J. Leggett, \textit{Modern Trends in the
Theory of Condensed Matter} (Springer-Verlag, Berlin, 1980), pp. 13-27.

\bibitem{NSR1985}P. Nozières and S. Schmitt-Rink, J. Low Temp. Phys.
\textbf{59}, 195 (1985).

\bibitem{SadeMelo1993}C. A. R. Sá de Melo, M. Randeria, and J. R.
Engelbrecht, Phys. Rev. Lett. \textbf{71}, 3202 (1993).

\bibitem{Hu2006}H. Hu, X.-J. Liu, and P. D. Drummond, Europhys. Lett.
\textbf{74}, 574 (2006).

\bibitem{Engelbrecht1992}J. R. Engelbrecht and M. Randeria, Phys.
Rev. B \textbf{45}, 12419 (1992).

\bibitem{Shenoy2011}V. B. Shenoy and T.-L. Ho, Phys. Rev. Lett. \textbf{107},
210401 (2011).

\bibitem{note1}We refer to the Appendix of Ref. \cite{He2015} for
a detailed discussion.

\bibitem{Wild2012}R. J. Wild, P. Makotyn, J. M. Pino, E. A. Cornell,
and D. S. Jin Phys. Rev. Lett. \textbf{108}, 145305 (2012).

\bibitem{Hu2011NJP}H. Hu, X.-J. Liu, and P. D. Drummond, New J. Phys.
\textbf{13}, 035007 (2011).

\bibitem{Schakel2010}A.M.J. Schakel, arXiv:1007.3452 (2010).

\bibitem{Hu2008}H. Hu, X.-J. Liu, and P. D. Drummond, Phys. Rev.
A \textbf{77}, 061605(R) (2008).

\bibitem{Hu2010}H. Hu, X.-J. Liu, and P. D. Drummond, New J. Phys.
\textbf{12}, 063038 (2010).

\bibitem{Hu2011PRA}H. Hu, X.-J. Liu, and P. D. Drummond, Phys. Rev.
A \textbf{83}, 063610 (2011).

\bibitem{note2}It is easy to show that, in the unitary limit, the
second-order virial coefficients for an attractive and repulsive gas
differ by a minus sign.

\bibitem{Beth1937}E. Beth and G. E. Uhlenbeck, Physica \textbf{4},
915 (1937).

\bibitem{Hoinka2013}S. Hoinka, M. Lingham, K. Fenech, H. Hu, C. J.
Vale, J. E. Drut, and S. Gandolfi, Phys. Rev. Lett. \textbf{110},
055305 (2013).\end{thebibliography}
\end{document}